\documentclass[aps,12pt,prb]{revtex4}
\usepackage{enumerate}
\usepackage{amsmath}

\begin{document}
\baselineskip18pt

\title{Indeterminacy relations in random dynamics}
\author{Piotr Garbaczewski\thanks{electronic address pgar@uni.opole.pl}}
\affiliation{Institute of Physics, University of Opole, 45-052
Opole, Poland}
\begin{abstract}
We analyze  various   uncertainty  measures  for  spatial
diffusion processes. In this manifestly non-quantum setting, we
focus on the existence issue of  complementary pairs whose joint
dispersion measure has strictly  positive lower bound.
\end{abstract}
\maketitle \noindent {\bf Keywords:} Information functionals;
dynamics of probability densities; entropy dynamics; diffusion processes;
Smoluchowski processes;  position-momentum dispersion bounds
\vskip1.0cm

\section{Conceptual background}

We take inspiration from the classic  meta-theorem in harmonic
analysis\cite{folland}: " a non-zero function and its  Fourier
transform cannot both be sharply localized". This statement lies
at the core of standard quantum-mechanical position-momentum
indeterminacy relationship  in $L^2(R^n)$, but its  range of
validity extends to   time-frequency indeterminacy measures which are
 employed in the classical signal analysis:  Fourier transform is here  a key
 element.

We look for  a  fairly distant analogue of the above
"no simultaneous sharp localization"    statement
in the theory of spatial diffusion processes (Wiener process and
Smoluchowski processes as examples) whose density functions belong
to $L^1(R)$  and there is no   notion of momentum observable, nor
any    physically digestible notion of momentum.

Nonetheless, in this non-quantum setting,  we shall address the existence issue
for  pairs of complementary dispersion measures that mimic the
previously mentioned meta-relationship  and so preclude an
arbitrarily sharp localization for both. The peculiar point of our  analysis will
 be a disregard of any Fourier transform input.

 Let us   consider  continuous probability densities on the
real line, with or without an explicit  time-dependence: $\rho \in
L^1(R);  \int_R \rho (x) \, dx =1$.   Our minimal demand is  that
the  first and second moments  of
 each density are finite, so that we   can introduce a two-parameter family $\rho _{\alpha ,\sigma }(x)$,
 labeled by the mean value $ \langle x\rangle = \int x\, \rho (x)\,  dx=\alpha \in R $ and  the standard deviation
 (here, square root of the variance) $\sigma \in R^+ $, $\sigma ^2 = \langle (x- \langle x\rangle )^2\rangle $.

We assume that  $\rho (x) $ admits suitable  differentiability
properties and impose  the natural boundary data at finite
or infinite (employed in below) integration boundaries.

Let there be  given a one-parameter $\alpha $-family of densities
whose mean square deviation  value   is   fixed at  $\sigma $.  By
introducing the mean value:
\begin{equation}
\langle [\sigma ^2\nabla \ln \rho  + (x- \langle x\rangle
)]^2\rangle \geq 0
\end{equation}
we readily arrive at an inequality
\begin{equation}
{\cal{F}} (\rho ) \doteq  \langle ({\nabla \ln \rho })^2\rangle =
\int {\frac{(\nabla \rho )^2}{\rho }}  dx    \geq {\frac{1}{\sigma
^2}}  \label{inequality}
\end{equation}
in which a minimum  of ${\cal{F}}$ is achieved if  and only if
$\rho $ is  a $\sigma $-Gaussian, compare e.g.\cite{stam,furth}.
The functional ${\cal{F}} (\rho )$ is often named the Fisher information
 associated with  $\rho $.

The above Eq.~(\ref{inequality}) actually associates a primordial
"momentum-position"  indeterminacy relationship (here, devoid of
any quantum connotations)  with  the probability distributions
under consideration.  Namely, let $D$ be a positive diffusion
constant with dimensions of $\hbar  /2m$ or $k_BT/m\beta $,
 c. f.\cite{gar}.
 We define an  auxiliary (named  osmotic) velocity field
 $u=u(x)=  D\nabla \ln \rho $. There holds:
 \begin{equation}
 \Delta x \cdot \Delta u  \geq D     \label{brown}
 \end{equation}
which  correlates  the  position variance  $\Delta x=  \langle
[x-\langle x\rangle ]^2\rangle ^{1/2}$ with the  osmotic velocity
variance $\Delta u = \langle [u - \langle u \rangle ]^2\rangle
^{1/2} $. In the above, $mD$ may be interpreted as the lower bound
for the joint  position-momentum dispersion measure; at least on
formal grounds, $m \Delta u$ carries a dimension of a physical
momentum variable.

This property extends to time-dependent situations  and is known to
be  respected by diffusion-type processes\cite{viola}.  Its primary
version for the free Brownian motion  has been found by R.
F\"{u}rth\cite{furth}.

Given  $\rho (x)$ and a  suitable   function $f(x) $, we can readily
 generalize the  previous arguments.  Let us introduce notions of a variance and covariance (here,
 directly borrowed from the random
 variable analysis\cite{golin}) for $x$ and $f(x)$. By means of the Schwarz inequality, we get:
 \begin{equation}
  \langle [x- \langle x\rangle ]^2\rangle \cdot \langle [f - \langle f\rangle ]^2\rangle \geq
  (\langle [x- \langle x\rangle ]\cdot \langle [f - \langle f\rangle ]\rangle )^2  \, ,
  \end{equation}
  hence, accordingly
  \begin{equation}
 Var(x) \cdot  Var(f) \geq     Cov ^2(x,f) \, .
 \end{equation}
 We note that for an osmotic velocity field $u(x)$, we have $\langle u\rangle =0$ and $\langle x\cdot u\rangle = - D$. Therefore
 \begin{equation}
  Var (x)\cdot Var(u) \geq  Cov^2(x,u) = D^2 \, ,
  \end{equation}
 with a dispersion  bound  $D^2$,  as anticipated in Eq.~(\ref{brown}).

 The problem is that a  careful selection of a function $f(x)$ is necessary, if
    we   expect   $Cov ^2(x,f)$  to set a definite  (fixed) lower
    bound for the joint dispersion measure, like $Cov^2(x,u) = D^2 $ does in the above.

\section{Dynamics}

Let us consider  spatial diffusion processes  in one space
dimension, like e.g. standard  Smoluchowski processes and their
generalizations. Let there be given  $\dot{x} = b(x,t) +A(t)$ with
$\langle A(s)\rangle =0 \, , \, \langle A(s)A(s')\rangle =
\sqrt{2D} \delta (s-s')$ and the corresponding Fokker-Planck
equation for the probability density $\rho  \in L^1(R)$ which we
analyze under the natural boundary conditions:
\begin{equation}
\partial _t\rho =
D\triangle \rho -  \nabla \cdot ( b \rho ) \, .\label{fokker}
\end{equation}

We assume the gradient form for the forward drift  $b= b(x,t)$ and
take $D$ as a  diffusion   constant with dimensions of $k_BT/m\beta
$.
 By introducing $u(x,t) = D \nabla \ln \rho (x,t)$ we define the current velocity of the process
 $v(x,t) = b(x,t) - u(x,t)$, in terms of which the  continuity equation  $\partial _t \rho = - \nabla (v\rho )$ follows.
The diffusion current reads $j=v \rho $.

Automatically, we have an indeterminacy relationship
   $Var (x)\cdot Var(u) \geq  Cov^2(x,u) = D^2$.    The corresponding  inequality   for the current velocity field
$Var(x) \cdot  Var(v) \geq     Cov ^2(x,v)$,     does not involve
any obvious lower  bound.

 The  cumulative identity
 $Var (x)\cdot [Var(u) + Var(v)] \geq Cov^2(x,v) + D^2$, reproduced in Ref.\cite{golin}, does not convey any illuminating
  message about  the diffusion process. It  cannot be  directly  inferred
from the Fisher functional   ${\cal{F}}(\rho )$, if $\rho $ has a
non-quantum origin:   the major obstacle at this point  is that
there is no diffusive analogue of  the quantum momentum
observable.

Let us also mention  another   attempt\cite{shin}  to set an
uncertainty principle for general diffusion processes.
 If adopted to our convention
(natural boundary data), in view of $\langle u\rangle =0$ and $v= b-
u$, we have $\langle v\rangle = \langle b \rangle $.

For an arbitrary real constant $C\neq 0$, we obviously have:  $[C
\cdot ( v- \langle v\rangle ) + (x-\langle x\rangle ]^2 \geq 0$. The
mean value of this auxiliary inequality  reads:
\begin{equation}
C^2 (\Delta v) ^2  + 2C[\cdot  Cov(x,b) +   D] +  (\Delta x)^2 \geq
0\, .
\end{equation}
and is  non-negative for all $C$,  which  enforces a condition
\begin{equation}
[D + Cov(x,b)]^2 - (\Delta v) ^2\cdot (\Delta x)^2   \leq 0 \, .
\end{equation}
Note that  $Cov(x,v)= D + Cov(x,b)$, so we have in fact  an
alternative derivation  of the previous indeterminacy relationship
$Var(x) \cdot Var(v) \geq Cov^2(x,v)$    for the current velocity
field.  The problem of the existence
    (or not) of a lower  bound for the joint dispersion measure of
    $x$ and$v$  has been left untouched.

 This observation extends to standard  Smoluchowski processes, whose   forward drifts are
      proportional to externally imposed force fields, typically through $b =F/m\beta $. Therefore  the position-current
       velocity      dispersion correlation  is controlled by    $Cov(x,F)$.
  For  the free  Brownian motion (e.g. the Wiener process)
     we have $b=0$, and hence $Cov(x,v)=D$ is a genuine lower bound.

\section{Indeterminacy measures for diffusion processes}

We begin from a classic observation that, once we set  $b= -
2D\nabla \Phi $  with $\Phi = \Phi (x,t)$,  a substitution:
\begin{equation}
\rho (x,t) \doteq \theta _*(x,t) \exp [- \Phi (x,t)]
\end{equation}
with $\theta _*$ and $\Phi $ being real functions, converts the
Fokker-Planck equation Eq.~(\ref{fokker})
 into a generalized diffusion  equation for $\theta _*$:
 \begin{equation}
 \partial _t \theta _* = D \Delta \theta _* -  {\frac{{\cal{V}}(x,t)}{2mD}}\theta _*
 \end{equation}
 and its  formal time adjoint (admitting also a more familiar  interpretation of
 a consistency condition, i. e. of  an indirect definition of the potential ${\cal{V}}$)
\begin{equation}
\partial _t \theta = -D\Delta \theta  + {\frac{{\cal{V}}(x,t)}{2mD}}\theta
\end{equation}
  which is valid for  a real function $\theta (x,t) = \exp [- \Phi
  (x,t)]$.

  Accordingly, we deduce
\begin{equation}
{\frac{{\cal{V}}(x,t)}{2mD}} =  - \partial _t \Phi + {\frac{1}2}
({\frac{b^2}{2D}} + \nabla \cdot b) = - \partial _t \Phi +
D[(\nabla \Phi )^2 -\Delta \Phi ] \, . \label{fokkerpot}
\end{equation}
There holds  an obvious factorization property for the
Fokker-Planck probability density:
\begin{equation}
\rho (x,t) = \theta (x,t)  \cdot \theta _*(x,t)\, .
\end{equation}

{\bf Remark 1:} Notice an  affinity with  a  familiar quantum mechanical
factorization formula $\rho =  \psi ^*  \psi $,  albeit presently
realized in terms of two  real functions $\theta $ and $\theta
^*$, instead of a complex conjugate pair. This issue is elucidated
in some detail in the Appendix.\\

 We have:
\begin{equation}
{\cal{F}}(\rho ) = {\frac{1}{D^2}} \sigma ^2_u = \int dx (\theta
\theta ^*) [{\frac{\nabla \theta }{\theta }}  + {\frac{\nabla
{\theta _*}}{\theta _*}}]^2= \label{ha1}
\end{equation}
$$
4\int dx \nabla {\theta }_* {\nabla \theta }  + \int dx (\theta
\theta _*)[{\frac{\nabla \theta }{\theta }}  - {\frac{\nabla {\theta
_*}}{\theta _*}}]^2 \, .
$$

Since a continuity equation  $\partial _t \rho = - \nabla j$ is
identically fulfilled by
\begin{equation}
j(x,t) = \rho (x,t) v(x,t) = D(\theta _* \nabla \theta - \theta
\nabla \theta _*)
\end{equation}
we obviously get (after an integration by parts and accounting for  the generalized  diffusion equations):
\begin{equation}
{\cal{F}}(\rho )= 4 \int dx
(\nabla \theta )(\nabla \theta ^*)
  + {\frac{1}{D^2}} \langle v^2\rangle  =  - {\frac{4}D}\langle \partial _t\Phi \rangle   -{\frac{2}{mD^2}}
  \langle  {\cal{V}}\rangle + {\frac{1}{D^2}} \langle v^2\rangle  \, .  \label{fisher}
\end{equation}

The  $\langle \partial _t\Phi \rangle $ contributions
cancel away once we invoke  an explicit expression for
${\cal{V}}$. Therefore, we may  simplify our  further  discussion by
assuming that $ \partial _t\Phi =0$ identically. This amounts to
passing to  Smoluchowski diffusion processes.\\

{\bf Remark 2:}  Let us indicate a formal similarity  of  our
reasoning for diffusion-type processes to that  developed by J-C.
Zambrini\cite{zambrini,zambrini1} in the framework  of the
so-called Euclidean quantum mechanics. In the latter approach,
another (Euclidean)   version of the uncertainty relations  was
introduced, based on  accepting a skew-adjoint operator   $-
2mD\nabla $ as a Euclidean version of a  quantum  momentum
observable  $-i2mD\nabla $.   To establish a corresponding
Heisenberg indeterminacy relation one needs to   accept
   a hypothesis  that $ 0< \int dx (\nabla \theta )(\nabla \theta ^*)    =  - (1/2mD^2) {\cal{V}}  <\infty $ in Eq.~(\ref{fisher}).
    This property  clearly  is not respected by
  the free Brownian motion  where ${\cal{V}}=0$ and, in view of Eq.~(\ref{fokkerpot}), may surely be violated by
  drifted Brownian motions.\\

We have (compare e.g.\cite{gar}):
\begin{equation}
{\cal{F}}(\rho =\theta \theta _*)= {\frac{2}{mD^2}}  \langle
{\frac{mv^2}2} - {\cal{V}}\rangle \Rightarrow \langle {\frac{mv^2}2}
- {\frac{mu^2}2}- {\cal{V}}\rangle  =0  \, . \label{mean}
\end{equation}
The variances of osmotic and current velocity fields are correlated
as follows
\begin{equation}
\rho =\theta \theta _* \Longrightarrow  m^2[ (\Delta u)^2   -
(\Delta v)^2] = 2m [ {\frac{m\langle v\rangle ^2}2} - \langle
{\cal{V}}\rangle ]   \, .      \label{diff}
\end{equation}
 On the left-hand-side of Eq.~(\ref{diff}), there appears a
difference of variances for the current  and osmotic velocity
fields. This expression is not necessarily positive definite, unless $\langle
{\cal{V}}\rangle \leq 0$ for all times.

Let us make a guess that  $\Delta u > \Delta v$, in the least locally in time (in a  finite time interval).
Then, the resulting expression
\begin{equation}
 m^2(\Delta u)^2= m^2 \langle u^2\rangle =
  2m \langle {\frac{m v^2}2} - {\cal{V}}\rangle \doteq  (\Delta p_{u})^2 \geq {\frac{m^2D^2}{\sigma ^2}} \, ,
  \end{equation}
as we already know,  yields a dimensionally acceptable
position-momentum  indeterminacy relationship for  diffusion-type
processes,
\begin{equation}
  \Delta x \cdot  \Delta p_{u} \geq mD\, , \label{one}
\end{equation}
 where   $ \Delta p_{u} >0 $ may  be interpreted as the   pertinent    "momentum dispersion" measure.
  For the free Brownian motion we have ${\cal{V}} =0$  and $v= -u$, hence   Eq.~(\ref{brown}) is recovered.

Upon making  an opposite guess i. e.  admit $\Delta v >\Delta u$  (again. at least locally in  time),  in view of
${\cal{F}} \geq 1/\sigma ^2$, we  would have
\begin{equation}
 m^2 (\Delta v)^2  =
  m^2 (\Delta u)^2 +  2m [  \langle
{\cal{V}}\rangle   - {\frac{m\langle v\rangle ^2}2}]  \doteq (\Delta p_v)^2 \geq {\frac{m^2D^2}{\sigma ^2}}
\end{equation}
and thus
\begin{equation}
 \Delta x \cdot \Delta p_v\geq mD  \, \label{two}
 \end{equation}
 would ultimately arise.

The above  two  indeterminacy options (\ref{one})  and  (\ref{two}) are a consequence of
 a possibly indefinite  sign  for a difference
$\Delta u - \Delta v$ of standard deviations, in the course of a diffusion process.
This sign issue seems  to   be a local in time property and may  not persist in the asymptotic (large time) regime.
We shall give an argument towards a
non-existence of a fixed positive lower bound for the joint
 position-current velocity uncertainty
measure in the vicinity of an asymptotic stationary solution of  the involved   Fokker-Planck equation.

The   diffusive potential ${\cal{V}}$ is not quite
arbitrary  and has a pre-determined functional form,
Eq.~(\ref{fokkerpot}). Our general restriction on ${\cal{V}}$  is
that it should be a continuous and bounded from below function. In
the diffusive case this demand guarantees that $\exp(-tH)$ with
$H\doteq -D\Delta + (1/2mD){\cal{V}}$  is a legitimate dynamical
semigroup operator, such that  $\theta _*(t=0)= \theta _{*0} \rightarrow \theta
_*(t)=\exp(-Ht) \theta _{*0}$.  That would suggest that we may  expect $\langle
{\cal{V}}\rangle  \geq 0$, which however typically is not the case.

 We cannot dwell on the general issue of entropy methods
in the study of the large time asymptotic for solutions of the
diffusion equations. In case of Smoluchowski diffusion processes
we may take for granted that they asymptotically
approach\cite{mackey,gar} unique stationary solutions, for which
the current velocity $v$ identically vanishes. Then $\Delta v =0$
as well, while   $0< Var(x) < \infty $ ( e. g.  $\Delta x $ stays
finite).

In view of Eq. (\ref{mean}), an asymptotic value of  the  strictly positive Fisher functional
 ${\cal{F}}$ equals $-(2/mD^2) \langle {\cal{V}}\rangle
>0$.  Accordingly, to secure ${\cal{F}}>0$,  an expectation value of ${\cal{V}}$ with respect
to the stationary probability density must be negative. Even, under an assumption that
 ${\cal{V}}$  is bounded from below.

Consequently,  in the large time asymptotic we surely have
$(\Delta u)^2 >  (1/\sigma ^2)>   (\Delta v)^2$ and  $\Delta v
\rightarrow 0$, while  $\sigma $ has a finite limiting value (an exception  is  the free Brownian motion when $\sigma $ diverges).
 The validity of the above argument can be checked
by inspection, after invoking an explicit solution for the
Ornstein-Uhlenbeck process\cite{gar,mackey}.

 Thus, $\Delta x \cdot \Delta p_v\geq mD$ does not hold true in the
vicinity of the asymptotic solution. On the contrary, $\Delta x \cdot \Delta p_u\geq mD$ is universally valid.

\section{Appendix:  Quantum indeterminacy}

Given  an $L^2(R)$-normalized function  $\psi (x)$. We denote
$({\cal{F}}\psi )(p)$ its Fourier transform. The corresponding
probability densities follow:
 $\rho (x) = |\psi (x)|^2$ and $\tilde{\rho }(p) = |({\cal{F}}\psi )(p)|^2$.\\
We introduce the related position and momentum  information
(differential, e.g. Shannon) entropies:
\begin{equation}
{\cal{S}}(\rho )\doteq S_q =  - \langle \ln \rho \rangle = - \int
\rho (x) \ln \rho (x)  dx
\end{equation}
    and
    \begin{equation}
    {\cal{S}}(\tilde{\rho })\doteq
S_p= - \langle  \ln \tilde{\rho }\rangle =   - \int \tilde{\rho
}(p) \ln \tilde{\rho }(p)  dp
\end{equation}
 where  ${\cal{S}}$ denotes the  Shannon
 entropy for a continuous probability  distribution.  For the sake of clarity, we
 use dimensionless quantities, although there exists a consistent procedure for handling
  dimensional quantities in the Shannon entropy definition.

We assume both entropies to take finite values. Then, there holds
the familiar   entropic uncertainty relation:
\begin{equation}
S_q + S_p \geq  (1 + \ln \pi ) \, .  \label{uncertainty}
\end{equation}

If  following  conventions we define the squared  standard
deviation   value  for an observable $A$ in a pure state $\psi $
as $(\Delta A)^2 = (\psi , [A - \langle A\rangle ]^2 \psi )$ with
$\langle A \rangle  = (\psi , A\psi)$, then for the  position $X$
and momentum $P$ operators we have the following version of the
entropic uncertainty relation  (here expressed through so-called
entropy powers, $\hbar \equiv 1$):
\begin{equation}
\Delta X \cdot \Delta P \geq  {\frac{1}{2\pi e}} \,
 \exp[{\cal{S}}(\rho )  + {\cal{S}}(\tilde{\rho })] \geq {\frac{1}{2}} \,  \label{un}
\end{equation}
which is  an alternative version of the entropic uncertainty
relation.
  For Gaussian densities,   $ (2\pi e )\Delta X \cdot \Delta P =
 \exp[{\cal{S}}(\rho )  + {\cal{S}}(\tilde{\rho })]$ holds true, but the     minimum $1/2$
  on the right-hand-side of Eq.~(\ref{un}),   is  not necessarily  reached.

 Let us notice that in view of properties of  the Fourier transform, there is a complete symmetry between  the
 inferred  information-theory
 functionals. After the Fourier transformation,  taking into   account
 the entropic uncertainty relation Eq.~(\ref{uncertainty}), we arrive at\cite{mycielski}:
\begin{equation}
4 \tilde{\sigma }^2 \geq  2(e\pi )^{-1}  \exp[-2  \langle \ln
\tilde{\rho }\rangle ] \geq  (2e\pi ) \exp[ 2 \langle \ln  \rho
\rangle ] \geq \sigma ^{-2} \label{chain1}
\end{equation}

Let us consider a momentum operator $P$ that is conjugate  to the
position operator  $X$ in the adopted dimensional convention
$\hbar \equiv 1$. Setting $P= - i d/dx$ and presuming that all
averages are finite,  we get:
\begin{equation}
[\langle P^2\rangle - \langle P\rangle ^2] =  (\Delta P)^2=
\tilde{\sigma }^2 \, .
\end{equation}
The  standard indeterminacy relationship
 $\sigma  \cdot \tilde{\sigma }\geq (1/2)$  follows.

  In the above, no explicit time-dependence
has been  indicated, but all derivations go through with any
wave-packet  solution $\psi (x,t)$ of the Schr\"{o}dinger
equation. The  induced dynamics of
 probability densities may imply the  time-evolution of  entropies: $S_q(t), S_p(t)$ and thence the dynamics of
 quantum  uncertainty measures  $ \Delta X  (t) =  \sigma (t)$ and $ \Delta P(t)= \tilde{\sigma }(t) $.

We consider the Schr\"{o}dinger equation in  the form:
\begin{equation}
i\partial _t  \psi  = - D  \Delta  \psi   +
{\frac{{\cal{V}}}{2mD}} \label{Schroedinger} \psi \, .
\end{equation}
where the  potential ${\cal{V}}= {\cal{V}}(\overrightarrow{x},t)$
(possibly time-dependent)  is  a  continuous (it is useful, if
bounded from below)  function  with dimensions of
 energy, $D=\hbar /2m$.

By employing the  Madelung decomposition:
\begin{equation}
 \psi = \rho ^{1/2} \exp(is/2D) \, ,
 \end{equation}
   with the
phase function $s=s(x,t)$  defining  a (current)  velocity field
$v=\nabla s $,  we readily  arrive at the continuity equation
\begin{equation}
\partial _t \rho = - \nabla (v \rho )
\end{equation}
 and the generalized Hamilton-Jacobi equation:
\begin{equation}
\partial _ts +\frac{1}2 ({\nabla }s)^2 + (\Omega -  Q) = 0 \label{jacobi1}
\end{equation}
where  $\Omega = {\cal{V}}/m$ and,   after  introducing an
 osmotic velocity field $
u(x,t) = D\nabla \ln \rho (x,t)$ we have, compare e.g. our
discussion of Section I:
\begin{equation}
Q  =  2D^{2}{\frac{\Delta \rho ^{1/2}} {\rho ^{1/2}}} =
{\frac{1}2} u^2 + D \nabla \cdot u \, .
\end{equation}

If  a quantum mechanical expectation value of the standard
Schr\"{o}dinger  Hamiltonian  $\hat{H}= -(\hbar ^2/2m) \Delta + V$
  exists (i.e.  is finite\cite{gar2}),
  \begin{equation}
 \langle \psi | \hat{H}|\psi \rangle \doteq E < \infty
 \end{equation}
  then the unitary  quantum dynamics warrants that this value is a  constant
 of the Schr\"{o}dinger picture evolution:
\begin{equation}
{\cal{H}} = {\frac{1}2} [\left< {v}^2\right> + \left<
{u}^2\right>]  + \left<\Omega \right>  =  - \left< \partial _t
s\right>  \doteq {\cal{E}} = {\frac{E}m}= const  \, .
\label{total}
\end{equation}
Let  us notice that  $\langle u^2\rangle = - D \langle \nabla u
\rangle $ and therefore:
\begin{equation}
{\frac{D^2}{2}} {\cal{F}} =   {\frac{D^2}{2}} \int {\frac{1}{\rho
}} \left({\frac{\partial \rho }{\partial x }} \right)^2\, dx  =
\int \rho \cdot  {\frac{u^2}{2}} dx = -   \langle Q \rangle \, .
\label{Fisher1}
  \end{equation}

We  observe that  $D^2{\cal{F}}$  stands for  the  mean square
deviation  value  of a function $u(x,t)$ about its mean value
$\langle u \rangle =0$, whose vanishing is a consequence of the
boundary conditions (here, at infinity):
\begin{equation}
(\Delta u)^2 \doteq \sigma _u^2 =  \langle [u- \langle u\rangle
]^2\rangle = \langle u^2\rangle  = D^2 {\cal{F}} \, .
\end{equation}
The  mean square deviation of $v(x,t)$ about its mean value
$\langle v\rangle$ reads:
\begin{equation}
(\Delta v)^2 \doteq \sigma ^2_v = \langle v^2 \rangle - \langle v
\rangle ^2\, .
\end{equation}
It is clear, that with the definition $P= -i(2mD) d/dx$, the mean
value of the operator $P$ is related to the mean value of a
function $v(x,t)$  (we do not discriminate between  technically
different implementations  of the mean):  $\langle P\rangle =
m\langle v \rangle $. Accordingly,
\begin{equation}
\tilde{\sigma }^2  = (\Delta P)^2 = \langle P^2\rangle - \langle
P\rangle ^2
\end{equation}

Moreover, we can directly check that with $\rho = |\psi |^2$ there
holds\cite{hall}:
\begin{equation}
{\cal{F}}(\rho ) = {\frac{1}{D^2}} \sigma ^2_u = \int dx |\psi
|^2[\psi '(x)/\psi (x) + {\psi ^*}'(x) /\psi ^*(x)]^2=
\label{ha}
\end{equation}
$$
4\int dx {\psi ^*}'(x) {\psi '}(x)  + \int dx |\psi (x)|^2 [ \psi
'(x)/\psi (x) -{\psi ^*}'(x) /\psi ^*(x)]^2  =
$$
$$
{\frac{1}{m^2D^2}} [ \langle P^2\rangle - m^2 \langle v^2 \rangle
] = {\frac{1}{m^2D^2}}[ (\Delta P)^2 -  m^2 \sigma ^2_v]
$$
i.e.
\begin{equation}
 m^2(\sigma ^2_u + \sigma ^2_v) = \tilde{\sigma }^2 \, . \label{heisenberg}
\end{equation}
  It is interesting to notice that $\langle (P - mv )\rangle  =0$
and the corresponding mean square deviation  reads: $
 \langle (P-mv)^2\rangle = \langle P^2\rangle - m^2\langle v^2\rangle = m^2D^2
 {\cal{F}}$.

By passing to dimensionless quantities in Eqs.~(\ref{ha}) (e.g.
$2mD\equiv 1$), and denoting $p_{cl} \doteq
 (\arg \,  \psi (x,t) )' $ we get:
\begin{equation}
{\cal{F}} = 4[\langle P^2\rangle - \langle p^2_{cl}\rangle ] =
4[(\Delta P)^2 -  (\Delta p_{cl})^2] = 4[\tilde{\sigma }^2 -
\tilde{\sigma }_{cl}^2]
\end{equation}
and therefore:
 \begin{equation}
4\tilde {\sigma }^2 \geq 4[\tilde{\sigma }^2 - \tilde{\sigma
}_{cl}^2] = {\cal{F}}      \geq (2\pi e)  \exp [-2{\cal{S}}(\rho
)] \geq  {\frac{1}{\sigma ^2}} \, .
\end{equation}
We recall that all "tilde" quantities can be  deduced from the
once  given $\psi $ and  its Fourier transform $\tilde{\psi}$.

   As  a side comment let us add that a direct consequence of the  mean  energy conservation law Eq.~(\ref{total}) are identities:
   $\langle P^2\rangle /2m = E - \langle {\cal{V}}\rangle $
   and
    \begin{equation}
    {\cal{F}} =  {\frac{1}{m^2D^2}} [ \langle P^2\rangle - m^2 \langle v^2 \rangle ] =
    {\frac{1}{D^2}} [2({\cal{E}} - \langle \Omega \rangle ) - \langle v^2 \rangle ]
\end{equation}
plus a complementary expression for the variance of the momentum
observable:
\begin{equation}
  (\Delta P)^2= 2 m  (E - \langle [{\frac{m}2}  \langle v \rangle ^2  + {\cal{V}}]\rangle
  )\, .
  \end{equation}

\end{document}